\documentclass[10pt]{article}
\usepackage{graphicx}
\usepackage{amsmath}
\usepackage{amssymb}
\usepackage{caption2}
\usepackage[figuresright]{rotating}
\begin{document}
\bibliographystyle{prsty}
\begin{center}
{\large {\bf \sc{  Isospin eigenstates of the color singlet-singlet type pentaquark states }}} \\[2mm]
Xiu-Wu Wang$^{*\dagger}$\footnote{E-mail: wangxiuwu2020@163.com.  },
Zhi-Gang  Wang$^*$\footnote{E-mail: zgwang@aliyun.com.  },
Guo-Liang Yu$^*$
and Qi Xin$^{*\dagger}$ \\
 Department of Physics, North China Electric Power University, Baoding 071003, P. R. China$^*$\\
 School of Nuclear Science and Engineering, North China Electric Power University, Beijing 102206, P. R. China$^\dagger$
\end{center}

\begin{abstract}
In this article, we construct the color singlet-singlet type five-quark currents with the isospins $(I,I_3)=(\frac{1}{2},\frac{1}{2})$ and $(\frac{3}{2},\frac{1}{2})$ unambiguously to  explore the $\bar{D}\Sigma_c$, $\bar{D}\Sigma_c^*$, $\bar{D}^*\Sigma_c$ and $\bar{D}^*\Sigma_c^*$ pentaquark states  via  the QCD sum rules for the first time, where the $\bar{D}$, $\Sigma_c$, $\cdots$ represent the color-singlet clusters having the same quantum numbers as the corresponding physical mesons or baryons. The numerical results support assigning the $P_c(4312)$, $P_c(4380)$, $P_c(4440)$ and $P_c(4457)$ as   the $\bar{D}\Sigma_c$, $\bar{D}\Sigma_c^*$, $\bar{D}^*\Sigma_c$ and $\bar{D}^*\Sigma_c^*$ pentaquark  states with the isospin $I=\frac{1}{2}$, respectively. The corresponding $\bar{D}\Sigma_c$, $\bar{D}\Sigma_c^*$, $\bar{D}^*\Sigma_c$ and $\bar{D}^*\Sigma_c^*$ pentaquark  states with the isospin $I=\frac{3}{2}$ have slightly larger masses, the observations of the higher pentaquark  candidates in the $J/\psi \Delta$ invariant mass spectrum would  shed light on the nature of the $P_c$ states, and make contributions in distinguishing  the scenarios of color antitriplet-antitriplet-antitriplet type and color singlet-singlet type pentaquark  states.
\end{abstract}



\section{Introduction}
  In 1964, M. Gell-mann proposed that the multiquark states could exist \cite{Gellmann}. Theoretically, there is no forbiddance for the existence of the exotic states which   cannot be embedded  into the conventional charmonium spectrum. Since the observation of the  $X(3872)$ by the Belle collaboration in 2003 \cite{Choi}, many   exotic $X$, $Y$, $Z$ particles have been observed at the Belle, BaBar, BESIII and LHCb collaborations  \cite{PDG}. The masses of some exotic states are close to the known two-particle thresholds, and lead to the possible hadronic molecule interpretations \cite{Guo1}, namely, the bound states of the meson-meson, baryon-meson or baryon-baryon. In 2015, the LHCb collaboration observed two pentaquark candidates $P_c(4380)$ and $P_c(4450)$ via  analysis of the $\Lambda_b^0\rightarrow J/\psi K^-p$ decays \cite{RAaij1}. In 2019, the LHCb collaboration re-investigated the experimental  data with
order of magnitude larger than that previously analyzed by the LHCb collaboration, and observed a narrow pentaquark candidate $P_c(4312)$ in the $J/\psi p$ mass spectrum \cite{RAaij2}, and proved that the $P_c(4450)$ consists  of two narrow overlapping peaks  $P_c(4440)$ and $P_c(4457)$. The measured Breit-Wigner masses and widths of the four exotic structures are
\begin{eqnarray}
\notag && P_c(4312):M=4311.9\pm0.7^{+6.8}_{-0.6}\,\rm{MeV}\, ,\,\,\,\Gamma=9.8\pm2.7^{+3.7}_{-4.5}\,\rm{MeV}\,,\\
\notag && P_c(4380):M=4380\pm8\pm29\,\rm{MeV}\, ,\,\, \,\Gamma=205\pm18\pm86\,\rm{MeV}\,,\\
\notag && P_c(4440):M=4440.3\pm1.3^{+4.1}_{-4.7}\,\rm{MeV}\,, \,\,\,\Gamma=20.6\pm4.9^{+8.7}_{-10.1}\,\rm{MeV}\,,\\
&& P_c(4457):M=4457.3\pm0.6^{+4.1}_{-1.7}\,\rm{MeV}\,, \,\,\,\Gamma=6.4\pm2.0^{+5.7}_{-1.9}\,\rm{MeV}\, ,
\end{eqnarray}
respectively \cite{RAaij1,RAaij2}.

Those  resonances  lie just a few $\rm{MeV}$ below the thresholds of the hidden-charm meson-baryon pairs $\bar{D}\Sigma_c$, $\bar{D}\Sigma_c^*$, $\bar{D}^*\Sigma_c$ and $\bar{D}^*\Sigma_c^*$, respectively. Now a typical interpretation of the $P_c(4312)$, $P_c(4380)$, $P_c(4440)$ and $P_c(4457)$ is that they are the S-wave hidden-charm meson-baryon molecules, and have definite isospin $I$, spin $J$ and parity $P$. For example, in Ref.\cite{Manuel}, it is proposed that the $P_c(4440)$ and $P_c(4457)$ are the $\bar{D}^*\Sigma_c$ bound states  with the $J^P=\frac{1}{2}^-$ and $\frac{3}{2}^-$ respectively via the one-pion exchange potential between the heavy antimeson and heavy baryon,  the result is consistent with the conclusion obtained  in Ref.\cite{MingZhu} via the one-boson-exchange model. Interestingly, the isospins are considered via  the one-pion/one-boson-exchange potential model  in Ref.\cite{FuLai} and a series of hidden-charm antimeson-baryon pentaquark molecules are predicted.  In Ref.\cite{MengLin}, the $\bar{D}^{(*)}\Sigma_c^{(*)}$ molecular states are studied via a coupled-channel formalism with the scattering potential involving both the  one-pion exchange and short-range operators constrained by the heavy quark spin symmetry, while in Ref.\cite{JunHe}, the $P_c(4440)$ and $P_c(4457)$ are interpreted as the $\bar{D}^*\Sigma_c$ bound states  with the $J^P=\frac{1}{2}^-$ and $\frac{3}{2}^-$, respectively  via the quasipotential Bethe-Salpeter equation approach.

Among the popular theoretical tools,  the QCD sum rules approach is a powerful theoretical tool in studying the exotic states, the $P_c(4312)$, $P_c(4380)$, $P_c(4440)$ and $P_c(4457)$ have been studied with the QCD sum rules, irrespective of being assigned as pentaquark states \cite{WZG-EPJC-1508,WZG-EPJC-1509,WZG-IJMPA-1905,WZG-IJMPA-2011} or pentaquark molecular states \cite{ChenHX-penta-mole,ChenHX-penta-mole-2,Azizi-pneta-mole,ZhangJR-penta-mole,WZG-penta-mole-1806,WZG-penta-mole-CPC,WZG-penta-mole-CPC-decay}. In the QCD sum rules, we choose the local five-quark currents, both the pentaquark states and  molecular states are compact objects, it is better to call the pentaquark molecular states as the
color singlet-singlet type pentaquark states, thereafter, we will prefer the nomenclature "color singlet-singlet type pentaquark states".

 If we prefer  interpretations of the  color singlet-singlet type pentaquark states and the theoretical approach of the QCD sum rules,  we should distinguish  their isospins and investigate their properties in an unambiguous way, however, in previous works, the isospins of the interpolating currents were not specified \cite{ChenHX-penta-mole,ChenHX-penta-mole-2,Azizi-pneta-mole,ZhangJR-penta-mole,WZG-penta-mole-1806,WZG-penta-mole-CPC,WZG-penta-mole-CPC-decay},
 the currents  couple potentially not only to the pentaquark  states with the isospin $I=\frac{1}{2}$ but also to the ones with the isospin $I=\frac{3}{2}$, there are unknown uncertainties.  Since those $P_c$ states were discovered in the $J/\psi p$ invariant mass spectrum, their isospins should be $I=\frac{1}{2}$ considering for conservation of the isospins in the strong interactions, and we should specify the isospins of the interpolating currents to make robust predictions,
  it is the key issue to solve the puzzle of those $P_c$ states.  In the present work, we explore the color singlet-singlet type pentaquark states with the $I=\frac{1}{2}$ and $\frac{3}{2}$ via the QCD sum rules in an systematic way.

The article is arranged as follows: we obtain the QCD sum rules for the color singlet-singlet type pentaquark states in Sect.2; we present the numerical results and discussions in Sect.3; Sect.4 is reserved for our conclusions.

\section{QCD sum rules for the color singlet-singlet type pentaquark states}
The $u$ quark and $d$ quark have the isospin $I=\frac{1}{2}$, in details, $\widehat{I}u=\frac{1}{2}u$ and $\widehat{I}d=-\frac{1}{2}d$, where the $\widehat{I}$ is the isospin operator. Then the $\bar{D}^0$, $\bar{D}^{*0}$, $\bar{D}^-$, $\bar{D}^{*-}$, $\Sigma_c^+$, $\Sigma_c^{*+}$, $\Sigma_c^{++}$ and $\Sigma_c^{*++}$ correspond to the eigenstates $|\frac{1}{2},\frac{1}{2}\rangle$, $|\frac{1}{2},\frac{1}{2}\rangle$, $|\frac{1}{2},-\frac{1}{2}\rangle$, $|\frac{1}{2},-\frac{1}{2}\rangle$, $|1,0\rangle$, $|1,0\rangle$, $|1,1\rangle$ and $|1,1\rangle$ in the isospin space $|I,I_3\rangle$, respectively.
And we can construct the following color-singlet currents to interpolate them,
\begin{eqnarray}
J^{\bar{D}^0}(x)&=&\bar{c}(x)i\gamma_5u(x)\, ,\nonumber \\
J^{\bar{D}^-}(x)&=&\bar{c}(x)i\gamma_5d(x)\, ,\nonumber \\
J^{\bar{D}^{*0}}_{\mu}(x)&=&\bar{c}(x)\gamma_{\mu} u(x)\, , \nonumber\\
J^{\bar{D}^{*-}}_{\mu}(x)&=&\bar{c}(x)\gamma_{\mu} d(x)\, , \nonumber\\
J^{\Sigma_c^+}(x)&=&\varepsilon^{ijk}u^{iT}(x)C\gamma_{\mu}d^j(x)\gamma^{\mu}\gamma_5c^k(x)\, ,\nonumber\\
J^{\Sigma_c^{++}}(x)&=&\varepsilon^{ijk}u^{iT}(x)C\gamma_{\mu}u^j(x)\gamma^{\mu}\gamma_5c^k(x)\, , \nonumber\\
J^{\Sigma_c^{*+}}_{\mu}(x)&=&\varepsilon^{ijk}u^{iT}(x)C\gamma_{\mu}d^j(x)c^k(x)\, , \nonumber\\
J^{\Sigma_c^{*++}}_{\mu}(x)&=&\varepsilon^{ijk}u^{iT}(x)C\gamma_{\mu}u^j(x)c^k(x)\, ,
\end{eqnarray}
the superscripts $i, j, k$ are color indices and the $C$ represents the charge conjugation matrix.
Accordingly, we can construct the color singlet-singlet type five-quark currents to interpolate the $\bar{D}^{(*)}\Sigma_c^{(*)}$-type pentaquark  sates,
where the $\bar{D}^{(*)}$ and $\Sigma_c^{(*)}$ represent the color-singlet clusters  having the same quantum numbers as the physical states  $\bar{D}^{(*)}$ and $\Sigma_c^{(*)}$, respectively,
and we write down the two-point correlation functions,
\begin{eqnarray}
 \Pi(p)&=&i\int d^4x e^{ip\cdot x}\langle 0 |T\left\{ J (x) \bar{J}(0) \right\}| 0\rangle \, ,\nonumber\\
\Pi_{\mu\nu}(p)&=&i\int d^4x e^{ip\cdot x}\langle 0 |T\left\{ J_{\mu} (x) \bar{J}_{\nu}(0) \right\}| 0\rangle \, ,\nonumber\\
\Pi_{\mu\nu\alpha\beta}(p)&=&i\int d^4x e^{ip\cdot x}\langle 0 |T\left\{ J_{\mu\nu} (x) \bar{J}_{\alpha\beta}(0) \right\}| 0\rangle \, ,
\end{eqnarray}
where the currents
\begin{eqnarray}
J(x)&=&J_{\frac{1}{2}}^{\bar{D}\Sigma_c}(x)\, , \,\,\,J_{\frac{3}{2}}^{\bar{D}\Sigma_c}(x)\, ,\nonumber\\
J_{\mu} (x)&=&J_{\frac{1}{2};\mu}^{\bar{D}\Sigma_c^*}(x)\, , \,\,\, J_{\frac{3}{2};\mu}^{\bar{D}\Sigma_c^*}(x)\, ,\,\,\, J_{\frac{1}{2};\mu}^{\bar{D}^{*}\Sigma_c}(x)\,, \,\,\,J_{\frac{3}{2};\mu}^{\bar{D}^{*}\Sigma_c}(x)\, , \nonumber\\
 J_{\mu\nu} (x)&=&J_{\frac{1}{2};\mu\nu}^{\bar{D}^{*}\Sigma_c^*}(x)\, , \, \,\, J_{\frac{3}{2};\mu\nu}^{\bar{D}^{*}\Sigma_c^*}(x)\, ,
  \end{eqnarray}
\begin{eqnarray}
J_{\frac{1}{2}}^{\bar{D}\Sigma_c}(x)&=&\frac{1}{\sqrt{3}}J^{\bar{D}^0}(x)J^{\Sigma_c^+}(x)-\sqrt{\frac{2}{3}}J^{\bar{D}^-}(x)J^{\Sigma_c^{++}}(x) \, , \nonumber\\
J_{\frac{3}{2}}^{\bar{D}\Sigma_c}(x)&=&\sqrt{\frac{2}{3}}J^{\bar{D}^0}(x)J^{\Sigma_c^+}(x)+\frac{1}{\sqrt{3}}J^{\bar{D}^-}(x)J^{\Sigma_c^{++}}(x)\, ,\nonumber\\
J_{\frac{1}{2};\mu}^{\bar{D}\Sigma_c^*}(x)&=&\frac{1}{\sqrt{3}}J^{\bar{D}^0}(x)J^{\Sigma_c^{*+}}_{\mu}(x)-\sqrt{\frac{2}{3}}J^{\bar{D}^-}(x)J^{\Sigma_c^{*++}}_{\mu}(x)\, ,\nonumber\\
J_{\frac{3}{2};\mu}^{\bar{D}\Sigma_c^*}(x)&=&\sqrt{\frac{2}{3}}J^{\bar{D}^0}(x)J^{\Sigma_c^{*+}}_{\mu}(x)+\frac{1}{\sqrt{3}}J^{\bar{D}^-}(x)J^{\Sigma_c^{*++}}_{\mu}(x)\, ,\nonumber\\
J_{\frac{1}{2};\mu}^{\bar{D}^{*}\Sigma_c}(x)&=&\frac{1}{\sqrt{3}}J^{\bar{D}^{*0}}_{\mu}(x)J^{\Sigma_c^+}(x)-\sqrt{\frac{2}{3}}J^{\bar{D}^{*-}}_{\mu}(x)J^{\Sigma_c^{++}}(x)\, ,\nonumber\\
J_{\frac{3}{2};\mu}^{\bar{D}^{*}\Sigma_c}(x)&=&=\sqrt{\frac{2}{3}}J^{\bar{D}^{*0}}_{\mu}(x)J^{\Sigma_c^+}(x)+\frac{1}{\sqrt{3}}J^{\bar{D}^{*-}}_{\mu}(x)J^{\Sigma_c^{++}}(x)\, ,\nonumber\\
J_{\frac{1}{2};\mu\nu}^{\bar{D}^{*}\Sigma_c^*}(x)&=&\frac{1}{\sqrt{3}}J^{\bar{D}^{*0}}_{\mu}(x)J^{\Sigma_c^{*+}}_{\nu}(x)-\sqrt{\frac{2}{3}}J^{\bar{D}^{*-}}_{\mu}(x)J^{\Sigma_c^{*++}}_{\nu}(x)+(\mu\leftrightarrow\nu)\, ,\nonumber\\
J_{\frac{3}{2};\mu\nu}^{\bar{D}^{*}\Sigma_c^*}(x)&=&\sqrt{\frac{2}{3}}J^{\bar{D}^{*0}}_{\mu}(x)J^{\Sigma_c^{*+}}_{\nu}(x)+\frac{1}{\sqrt{3}}J^{\bar{D}^{*-}}_{\mu}(x)J^{\Sigma_c^{*++}}_{\nu}(x)+(\mu\leftrightarrow\nu)\, ,
\end{eqnarray}
 the subscripts $\frac{1}{2}$ and $\frac{3}{2}$ represent the isospins $I$ \cite{WZG-penta-mole-CPC}.  The currents are the isospin eigenstates $|I,I_3\rangle=$
$|\frac{1}{2},\frac{1}{2}\rangle$ or $|\frac{3}{2},\frac{1}{2}\rangle$, respectively.

The currents $J(x)$, $J_\mu(x)$ and $J_{\mu\nu}(x)$ couple potentially  not only to the color singlet-singlet type  hidden-charm   pentaquark states   with  negative-parity but also to the ones with positive parity, we separate their  ground state contributions at the hadron side,
\begin{eqnarray}
  \Pi(p) & = & {\lambda^{-}_{\frac{1}{2}}}^2  {\!\not\!{p}+ M_{-} \over M_{-}^{2}-p^{2}  } +  {\lambda^{+}_{\frac{1}{2}}}^2  {\!\not\!{p}- M_{+} \over M_{+}^{2}-p^{2}  } +\cdots  \, ,\nonumber\\
  &=&\Pi_{\frac{1}{2}}^1(p^2)\!\not\!{p}+\Pi_{\frac{1}{2}}^0(p^2)\, ,
\end{eqnarray}
 \begin{eqnarray}
   \Pi_{\mu\nu}(p) & = & {\lambda^{-}_{\frac{3}{2}}}^2  {\!\not\!{p}+ M_{-} \over M_{-}^{2}-p^{2}  } \left(- g_{\mu\nu}\right)+  {\lambda^{+}_{\frac{3}{2}}}^2  {\!\not\!{p}- M_{+} \over M_{+}^{2}-p^{2}  } \left(- g_{\mu\nu}\right)   +\cdots  \, ,\nonumber\\
   &=&-\Pi_{\frac{3}{2}}^1(p^2)\!\not\!{p}\,g_{\mu\nu}-\Pi_{\frac{3}{2}}^0(p^2)\,g_{\mu\nu}+\cdots\, ,
\end{eqnarray}
 \begin{eqnarray}
\Pi_{\mu\nu\alpha\beta}(p) & = & {\lambda^{-}_{\frac{5}{2}}}^2  {\!\not\!{p}+ M_{-} \over M_{-}^{2}-p^{2}  } \left( g_{\mu\alpha}g_{\nu\beta}+g_{\mu\beta}g_{\nu\alpha}
\right)+ {\lambda^{+}_{\frac{5}{2}}}^2  {\!\not\!{p}- M_{+} \over M_{+}^{2}-p^{2}  } \left( g_{\mu\alpha}g_{\nu\beta}+g_{\mu\beta}g_{\nu\alpha}\right) +\cdots \, , \nonumber\\
&=&\Pi_{\frac{5}{2}}^1(p^2)\!\not\!{p}\left( g_{\mu\alpha}g_{\nu\beta}+g_{\mu\beta}g_{\nu\alpha}\right)+\Pi_{\frac{5}{2}}^0(p^2)\,\left( g_{\mu\alpha}g_{\nu\beta}+g_{\mu\beta}g_{\nu\alpha}\right)+ \cdots \, ,
\end{eqnarray}
where the subscripts $\frac{1}{2}$, $\frac{3}{2}$ and $\frac{5}{2}$ are the spins of the pentaquark  states, the subscripts/superscripts $\pm$ denote the positive-parity and
negative-parity, respectively, and we have smeared  the isospin indexes.  The pole residues are defined by
\begin{eqnarray}\label{J-lamda-1}
\langle 0| J (0)|P_{\frac{1}{2}}^{-}(p)\rangle &=&\lambda^{-}_{\frac{1}{2}} U^{-}(p,s) \, ,\nonumber  \\
\langle 0| J (0)|P_{\frac{1}{2}}^{+}(p)\rangle &=&\lambda^{+}_{\frac{1}{2}}i\gamma_5 U^{+}(p,s) \, ,
 \end{eqnarray}
\begin{eqnarray}\label{J-lamda-2}
\langle 0| J_{\mu} (0)|P_{\frac{3}{2}}^{-}(p)\rangle &=&\lambda^{-}_{\frac{3}{2}} U^{-}_\mu(p,s) \, , \nonumber \\
\langle 0| J_{\mu} (0)|P_{\frac{3}{2}}^{+}(p)\rangle &=&\lambda^{+}_{\frac{3}{2}}i\gamma_5 U^{+}_{\mu}(p,s) \, ,
\end{eqnarray}
 \begin{eqnarray}\label{J-lamda-3}
\langle 0| J_{\mu\nu} (0)|P_{\frac{5}{2}}^{-}(p)\rangle &=&\sqrt{2}\lambda^{-}_{\frac{5}{2}} U^{-}_{\mu\nu}(p,s) \, , \nonumber\\
\langle 0| J_{\mu\nu} (0)|P_{\frac{5}{2}}^{+}(p)\rangle &=&\sqrt{2}\lambda^{+}_{\frac{5}{2}}i\gamma_5 U^{+}_{\mu\nu}(p,s) \, ,
\end{eqnarray}
 where the $U^\pm(p,s)$, $U^{\pm}_\mu(p,s)$ and $U^{\pm}_{\mu\nu}(p,s)$ are the Dirac and  Rarita-Schwinger spinors, for all the technical details, one can consult Refs.\cite{WZG-EPJC-1508,WZG-EPJC-1509,WZG-IJMPA-1905,WZG-IJMPA-2011,WZG-penta-mole-1806,WZG-penta-mole-CPC}.

In the present work, we choose the components associated with the structures $\!\not\!{p}$,  $1$,   $\!\not\!{p}g_{\mu\nu}$, $g_{\mu\nu}$ and $\!\not\!{p}\left(g_{\mu\alpha}g_{\nu\beta}+g_{\mu\beta}g_{\nu\alpha}\right)$, $g_{\mu\alpha}g_{\nu\beta}+g_{\mu\beta}g_{\nu\alpha}$ in  the correlation functions $\Pi(p)$, $\Pi_{\mu\nu}(p)$ and $\Pi_{\mu\nu\alpha\beta}(p)$ respectively to investigate  the color singlet-singlet type pentaquark states with the spin-parity $J^P={\frac{1}{2}}^\mp$, ${\frac{3}{2}}^\mp$ and ${\frac{5}{2}}^\mp$, respectively.

We carry out the complex  operator product expansion, and analyze the contributions of  all kinds of vacuum condensates. Firstly, the contributions of the related vacuum condensates are tiny in the case of $k\geq\frac{3}{2}$ for the counting-rules in terms of the strong fine-structure constant  $\mathcal{O}(\alpha_s^k )$ \cite{Wang-tetraquark-QCDSR-1,Wang-tetraquark-QCDSR-2,Wang-molecule-QCDSR-1,Wang-molecule-QCDSR-2}, it is accurate enough for us to calculate the terms for $k\leq1$ \cite{wangxiuwu}. Secondly,  the highest dimension of the vacuum condensates is usually estimated from the leading order Feynman diagrams. In the present work, the correlation functions contain two heavy quark lines and three light quark lines. If each heavy quark line emits a gluon and each light quark line contributes a quark-antiquark pair, we obtain the quark-gluon operator $g_sG_{\alpha\beta}g_sG_{\eta\tau}\overline{q}q\overline{q}q\overline{q}q$ with the dimension $13$,
this operator can be factorized into the vacuum condensates $\langle\frac{\alpha_s}{\pi}GG\rangle\langle\overline{q}q\rangle^3$ and $\langle\overline{q}g_s\sigma Gq\rangle^2\langle\overline{q}q\rangle$. Thirdly, the four-quark condensates $\langle\overline{\psi}\psi\rangle^2=\sum_{u,s,d}\langle\overline{q}q\rangle^2$ are neglected as they come from condensations between the two heavy quark lines through equation of motion and  play a tiny role \cite{wangxiuwu}. Thus, in this work, there are solid reasons for us to choose the terms $\langle\bar{q}q\rangle$, $\langle\frac{\alpha_s}{\pi}GG\rangle$, $\langle\overline{q}g_s\sigma Gq\rangle$, $\langle\overline{q}q\rangle^2$, $\langle\frac{\alpha_s}{\pi}GG\rangle\langle\bar{q}q\rangle$, $\langle\overline{q}g_s\sigma Gq\rangle\langle\overline{q}q\rangle$, $\langle\bar{q}q\rangle^3$, $\langle\overline{q}g_s\sigma Gq\rangle^2$, $\langle\frac{\alpha_s}{\pi}GG\rangle\langle\overline{q}q\rangle^2$, $\langle\overline{q}g_s\sigma Gq\rangle\langle\bar{q}q\rangle^2$, $\langle\overline{q}q\rangle^4$, $\langle\overline{q}g_s\sigma Gq\rangle^2\langle\overline{q}q\rangle$ and $\langle\frac{\alpha_s}{\pi}GG\rangle\langle\overline{q}q\rangle^3$ in the  operator product expansions.

We  obtain the analytical spectral densities $\rho^1_{j,QCD}(s)$ and $\rho^0_{j,QCD}(s)$ at the quark-gluon level,  and take the
quark-hadron duality below the continuum thresholds  $s_0$ and introduce the weight functions $\sqrt{s}\exp\left(-\frac{s}{T^2}\right)$ and $\exp\left(-\frac{s}{T^2}\right)$ to obtain  the  QCD sum rules:
\begin{eqnarray}\label{QCDN}
2M_{-}{\lambda^{-}_{j}}^2\exp\left( -\frac{M_{-}^2}{T^2}\right)
&=& \int_{4m_c^2}^{s_0}ds \left[\sqrt{s}\rho^1_{j,QCD}(s)+\rho^0_{j,QCD}(s)\right]\exp\left( -\frac{s}{T^2}\right)\, ,
\end{eqnarray}
as we are only interested in the pentaquark  states with the negative parity, the explicit expressions of the  spectral densities $\rho^1_{j,QCD}(s)$ and $\rho^0_{j,QCD}(s)$ at the quark level  are neglected for simplicity.

We differentiate   Eq.\eqref{QCDN} with respect to  $\tau=\frac{1}{T^2}$, then eliminate the
 pole residues $\lambda^{-}_{j}$ with $j=\frac{1}{2}$, $\frac{3}{2}$, $\frac{5}{2}$ to  obtain the QCD sum rules for
 the masses of the color singlet-singlet type pentaquark states,
 \begin{eqnarray}\label{QCDSR-M}
 M^2_{-} &=& \frac{-\frac{d}{d \tau}\int_{4m_c^2}^{s_0}ds \,\left[\sqrt{s}\,\rho^1_{QCD}(s)+\,\rho^0_{QCD}(s)\right]\exp\left(- \tau s\right)}{\int_{4m_c^2}^{s_0}ds \left[\sqrt{s}\,\rho_{QCD}^1(s)+\,\rho^0_{QCD}(s)\right]\exp\left( -\tau s\right)}\, ,
 \end{eqnarray}
where the spectral densities $\rho_{QCD}^1(s)=\rho_{j,QCD}^1(s)$ and $\rho^0_{QCD}(s)=\rho^0_{j,QCD}(s)$.

\section{Numerical results and discussions}
We apply the standard values of the vacuum condensates $\langle\overline{q}q\rangle=-(0.24\pm0.01\;{\rm GeV})^3$, $\langle\overline{q}g_s\sigma Gq\rangle=m_0^2\langle\overline{q}q\rangle\;$GeV$^2$, $m_0^2=(0.8\pm0.1)\;{\rm GeV}^2$, $\langle\frac{\alpha_s}{\pi}GG\rangle=(0.33\;{\rm GeV})^4$ at the energy scale $\mu=1\;{\rm GeV}$ \cite{SVZ1,SVZ2,Reinders,ColangeloReview}, and choose the $
\overline{MS}$ mass $m_c(m_c)=1.275\pm0.025\;{\rm GeV}$ from the Particle Data Group \cite{PDG}. We consider the energy-scale dependence of those parameters,
\begin{eqnarray}
\notag \langle\overline{q}q\rangle(\mu)&&=\langle\overline{q}q\rangle(1{\rm GeV})\left[\frac{\alpha_s(1{\rm GeV})}{\alpha_s(\mu)}\right]^{\frac{12}{33-2n_f}}\, ,\\
\notag \langle\overline{q}g_s\sigma Gq\rangle(\mu)&& =\langle\overline{q}g_s\sigma Gq\rangle(1{\rm GeV})\left[\frac{\alpha_s(1{\rm GeV})}{\alpha_s(\mu)}\right]^{\frac{2}{33-2n_f}}\, ,\\
\notag  m_c(\mu)&&=m_c(m_c)\left[\frac{\alpha_s(\mu)}{\alpha_s(m_c)}\right]^{\frac{12}{33-2n_f}}\, ,\\
\notag \alpha_s(\mu)&&=\frac{1}{b_0t}\left[1-\frac{b_1}{b_0^2}\frac{\rm{log}\emph{t}}{t}+\frac{b_1^2(\rm{log}^2\emph{t}-\rm{log}\emph {t}-1)+\emph{b}_0\emph{b}_2}{b_0^4t^2}\right]\, ,
\end{eqnarray}
where $t=\rm{log}\frac{\mu^2}{\Lambda_{\emph{QCD}}^2}$, $\emph b_0=\frac{33-2\emph{n}_\emph{f}}{12\pi}$, $b_1=\frac{153-19n_f}{24\pi^2}$, $b_2=\frac{2857-\frac{5033}{9}n_f+\frac{325}{27}n_f^2}{128\pi^3}$
and $\Lambda_{QCD}=213$ MeV, $296$ MeV, $339$ MeV for the flavors $n_f=5,4,3$, respectively \cite{PDG,Narison}, In this paper, we choose the flavor number $n_f=4$ for all the pentaquark   states,  and apply the energy scale formula to determine the best energy scales of the QCD spectral densities \cite{WZG-EPJC-1508,WZG-IJMPA-1905,WZG-penta-mole-1806,WZG-penta-mole-CPC,Wang-tetraquark-QCDSR-1,Wang-tetraquark-QCDSR-2,Wang-molecule-QCDSR-1,Wang-molecule-QCDSR-2},
\begin{eqnarray}
\mu=\sqrt{M_{X/Y/Z/P}^2-4\mathbb{M}_c^2}\, ,
\end{eqnarray}
where the $\mathbb{M}_c$ is the effective charm quark mass, we choose the updated value $\mathbb{M}_c=1.85\pm0.01$ \rm GeV \cite{WZG-penta-mole-CPC}.

All the QCD sum rules should satisfy the pole dominance and convergence of the operator product expansion which are two basic criteria. What's more, we should obtain Borel platforms to avoid additional uncertainties originated from the Borel parameters. The selections of the suitable energy scales, continuum threshold parameters  and Borel parameters are accomplished  via trial and error: we tentatively choose  an energy scale  $\mu$ and a continuum threshold parameter $s_0$, then obtain the numerical value of the pentaquark  mass $M_P$ from  the QCD sum rules, and  judge whether or not the two basic criteria of the QCD sum rules (plus the constraint  $\sqrt{s_0}=M_P+0.6\sim 0.7\,\rm{GeV}$, plus the energy scale formula $\mu=\sqrt{M_{P}^2-4\mathbb{M}_c^2}$) are satisfied. If not, we choose another energy scale and another continuum threshold parameter until reach the satisfactory results.  In calculations, we define  the pole contributions (PC) as,
\begin{eqnarray}
{\rm PC}&=&\frac{\int_{4m_c^2}^{s_0}ds\left[\sqrt{s}\rho_{QCD}^1(s)+\rho_{QCD}^0(s)\right]\exp\left(-\frac{s}{T^2}\right)}
{\int_{4m_c^2}^{\infty}ds\left[\sqrt{s}\rho_{QCD}^1(s)+\rho_{QCD}^0(s)\right]\exp\left(-\frac{s}{T^2}\right)}\, .
\end{eqnarray}
 The convergence  of the operator product expansion is quantified via the contributions of the vacuum condensates of dimension $n$,
\begin{eqnarray}
D(n)&=&\frac{\int_{4m_c^2}^{s_0}ds\left[\sqrt{s}\rho_{QCD;n}^1(s)+\rho_{QCD;n}^0(s)\right]\exp\left(-\frac{s}{T^2}\right)}
{\int_{4m_c^2}^{s_0}ds\left[\sqrt{s}\rho_{QCD}^1(s)+\rho_{QCD}^0(s)\right]\exp\left(-\frac{s}{T^2}\right)}\, ,
\end{eqnarray}
 where the $\rho_{QCD;n}^1(s)$ and $\rho_{QCD;n}^0(s)$ represent the spectral densities with the vacuum condensates of dimension $n$ picked out from the $\rho_{QCD}^1(s)$ and $\rho_{QCD}^0(s)$, respectively, and the total contributions are normalized  to be 1.

 At last, we find  the best energy scales, the ideal continuum threshold parameters, the Borel windows, see Table \ref{BorelP-mass-residue},  the pole contributions for all the eight pentaquark states are around (or slightly larger than) $(40-60)\%$, thus, the pole dominance criterion for the QCD sum rules holds well.

\begin{figure}
 \centering
 \includegraphics[totalheight=5cm,width=7cm]{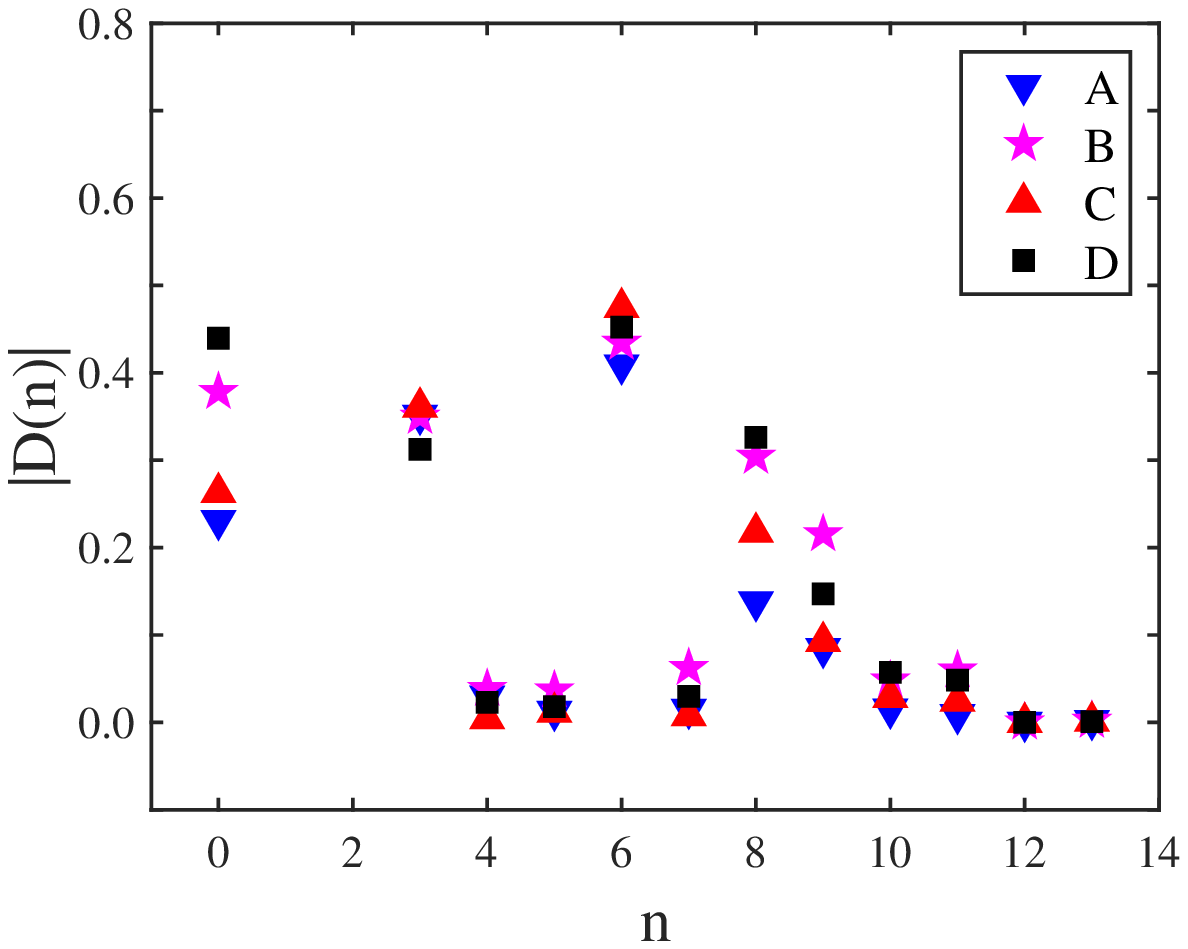}
 \includegraphics[totalheight=5cm,width=7cm]{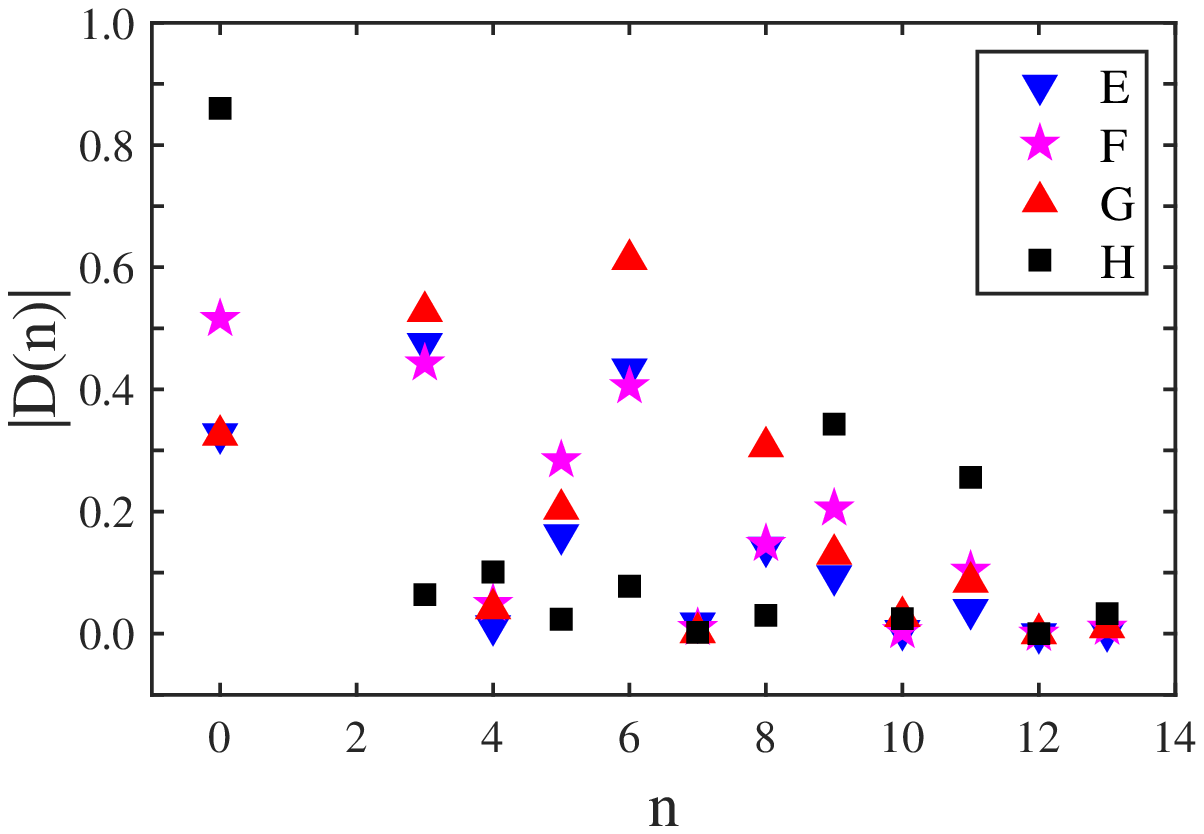}
 \caption{ The contributions of the vacuum condensates of dimension $n$,
where the A, B, C, D, E, F, G and H denote the pentaquarks $\bar{D}\Sigma_c$ with $I=\frac{1}{2}$, $\bar{D}\Sigma_c$ with
$I=\frac{3}{2}$, $\bar{D}\Sigma_c^*$ with $I=\frac{1}{2}$, $\bar{D}\Sigma_c^*$ with $I=\frac{3}{2}$, $\bar{D}^*\Sigma_c$ with
$I=\frac{1}{2}$, $\bar{D}^*\Sigma_c$ with $I=\frac{3}{2}$, $\bar{D}^*\Sigma_c^*$ with $I=\frac{1}{2}$ and
$\bar{D}^*\Sigma_c^*$ with $I=\frac{3}{2}$, respectively.}\label{OPE-fig}
\end{figure}

 The  absolute values of the normalized contributions $D(n)$ from the vacuum condensates are displayed in the Fig.\ref{OPE-fig}, where the highest dimensional condensate contributions $|D(12)|$ and $|D(13)|$ are approximately zero,  the most important contributions are mainly from the lowest order contributions $\langle \bar{q}q\rangle$, $\langle \bar{q}q\rangle^2$ and $\langle\bar qg_s\sigma Gq\rangle\langle\bar qq\rangle$, and the gluon condensate plays a less important role since $|D(4)|<5\%$ except  for the $\bar{D}^*\Sigma_c^*$ pentaquark  state  with the isospin $I=\frac{3}{2}$. All in all, the convergence of the operator expansions is very well satisfied.

\begin{figure}
 \centering
 \includegraphics[totalheight=5cm,width=7cm]{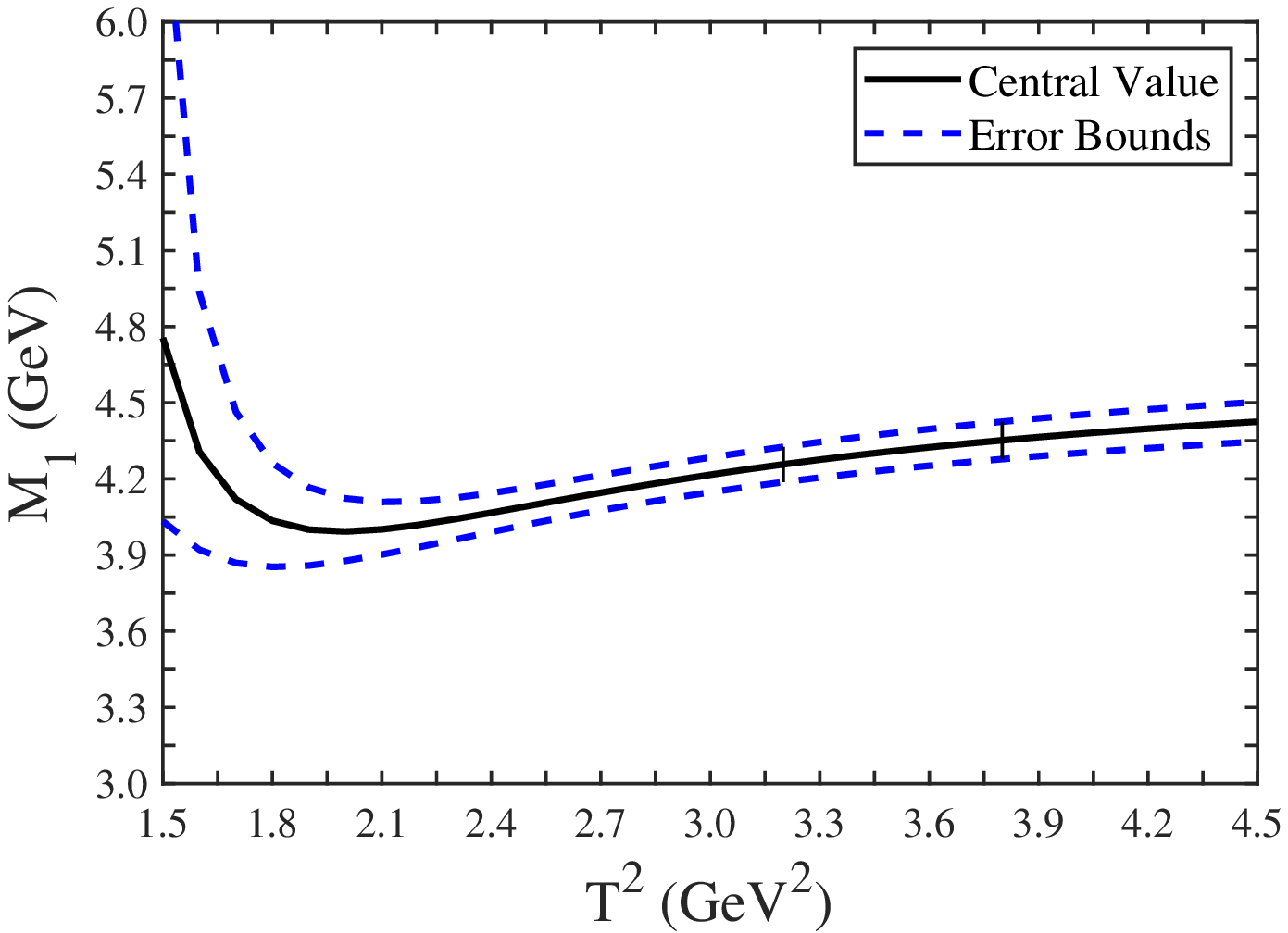}
 \includegraphics[totalheight=5cm,width=7cm]{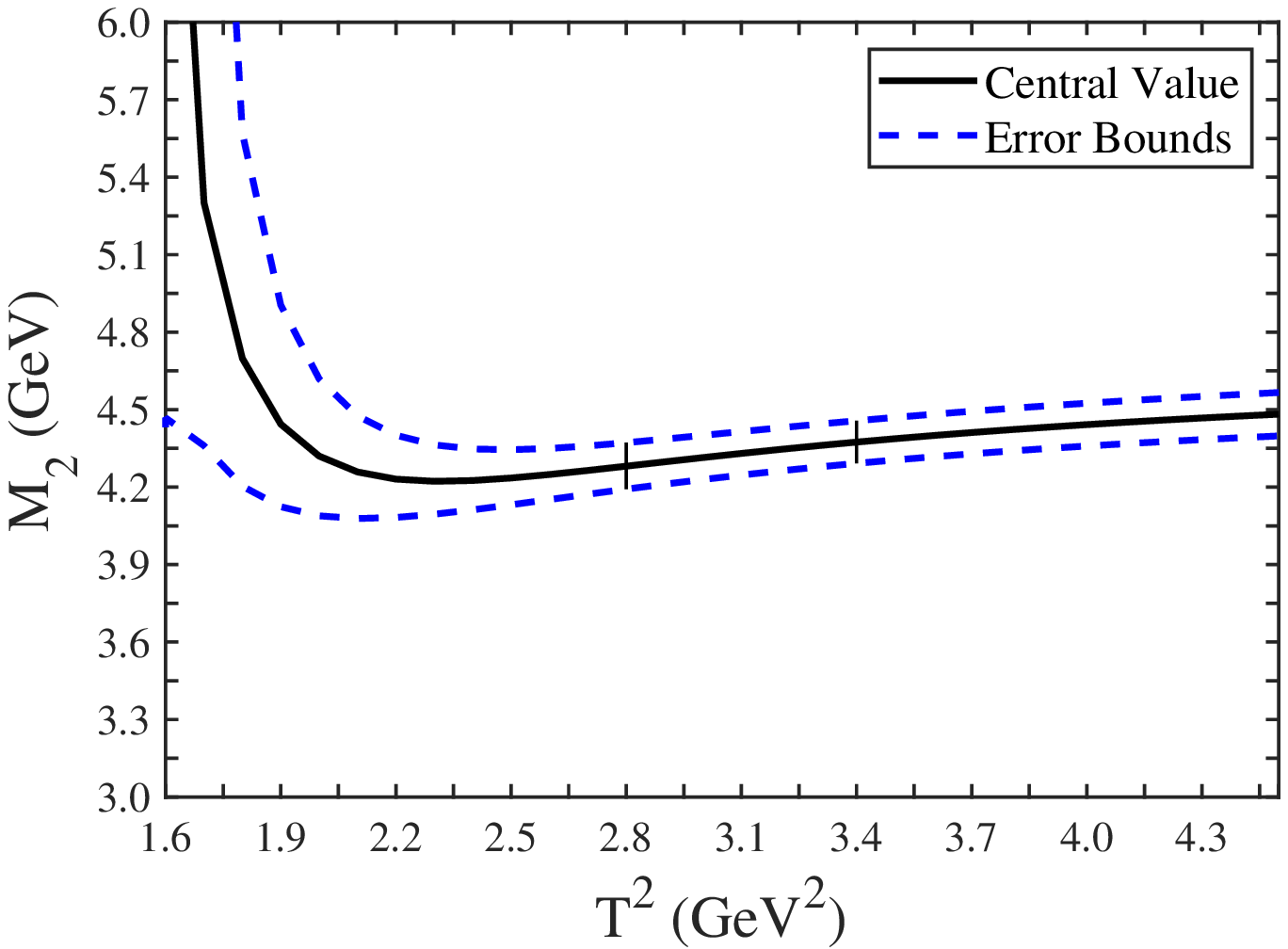}
 \includegraphics[totalheight=5cm,width=7cm]{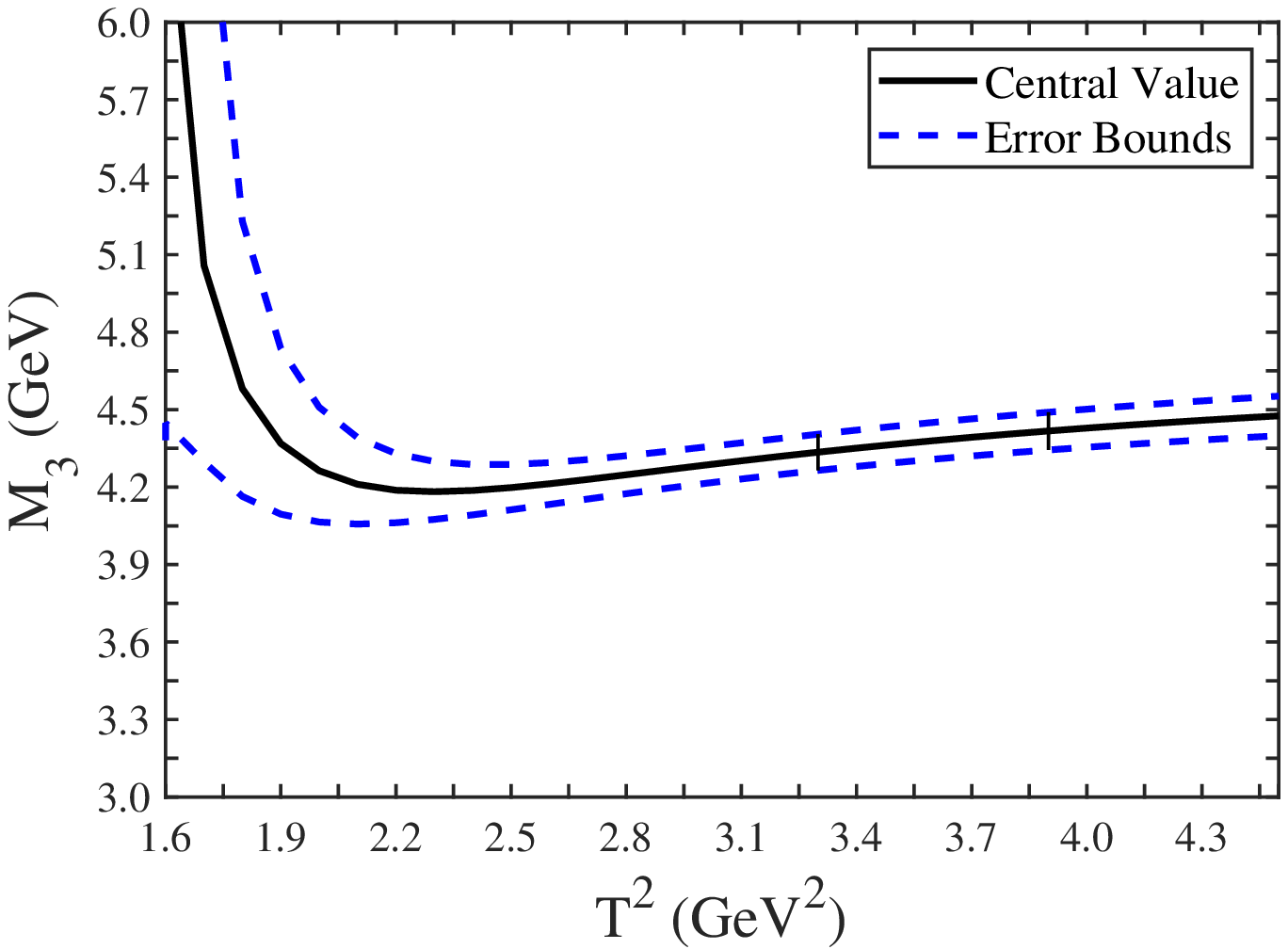}
 \includegraphics[totalheight=5cm,width=7cm]{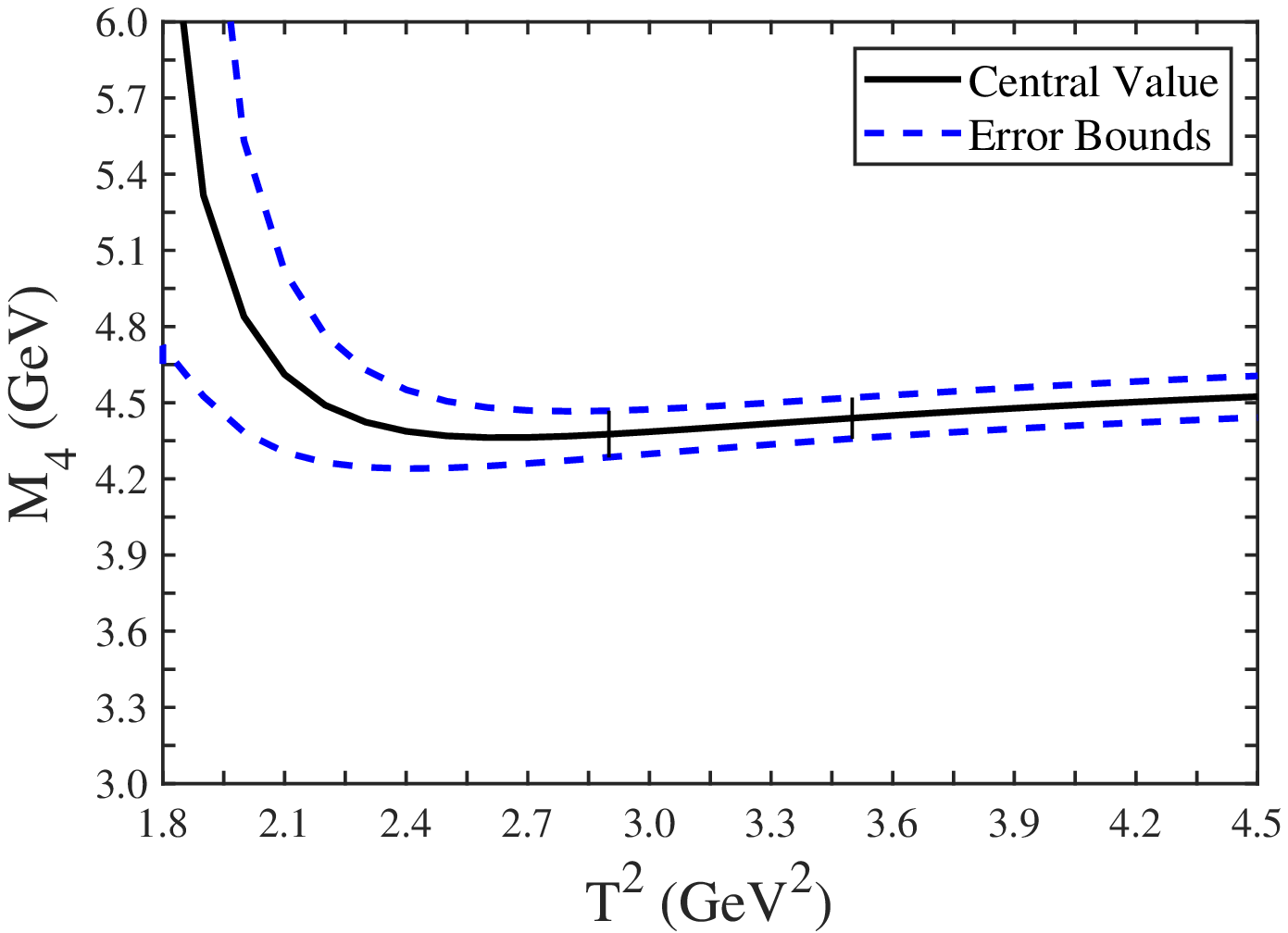}
 \includegraphics[totalheight=5cm,width=7cm]{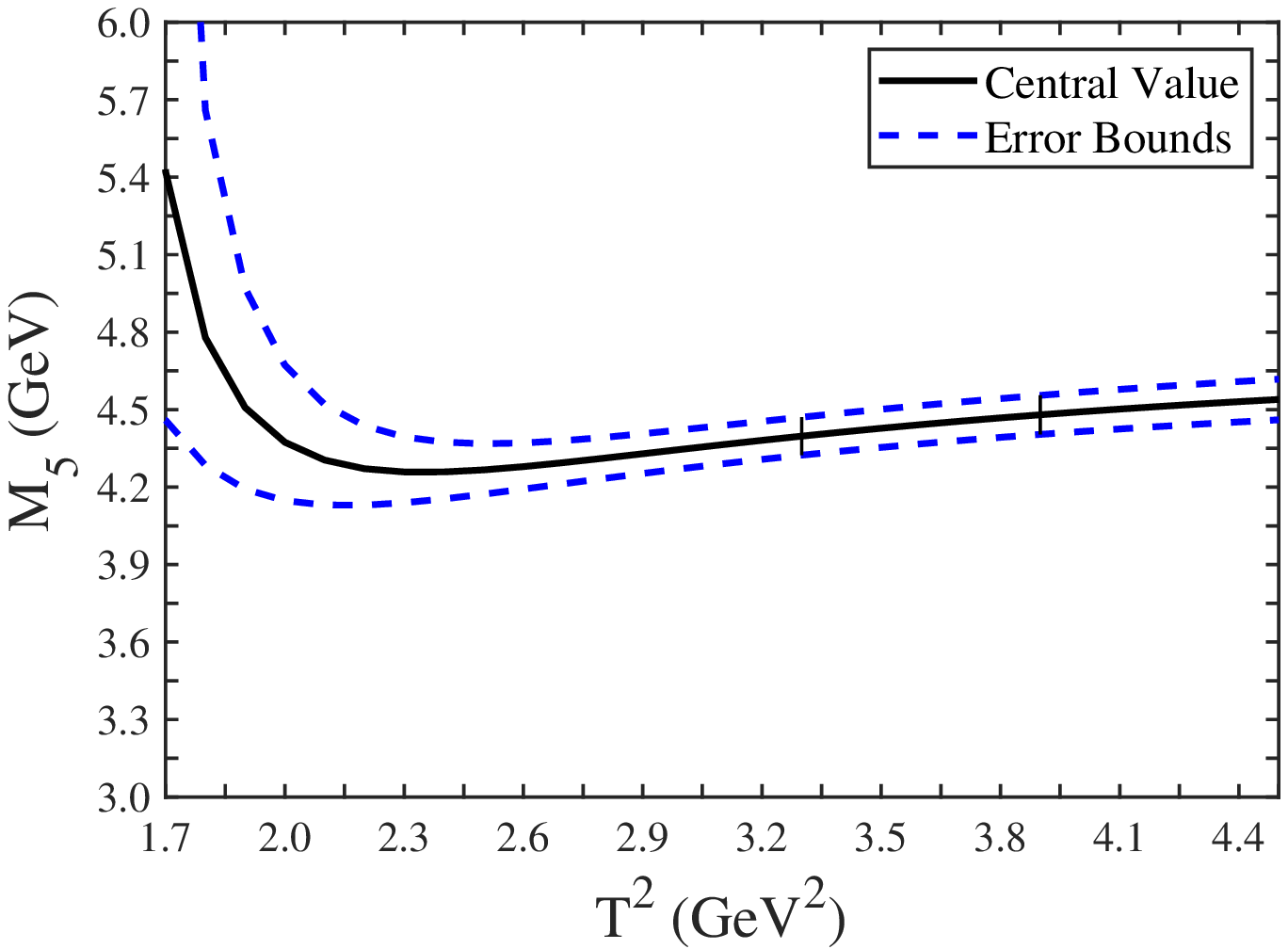}
 \includegraphics[totalheight=5cm,width=7cm]{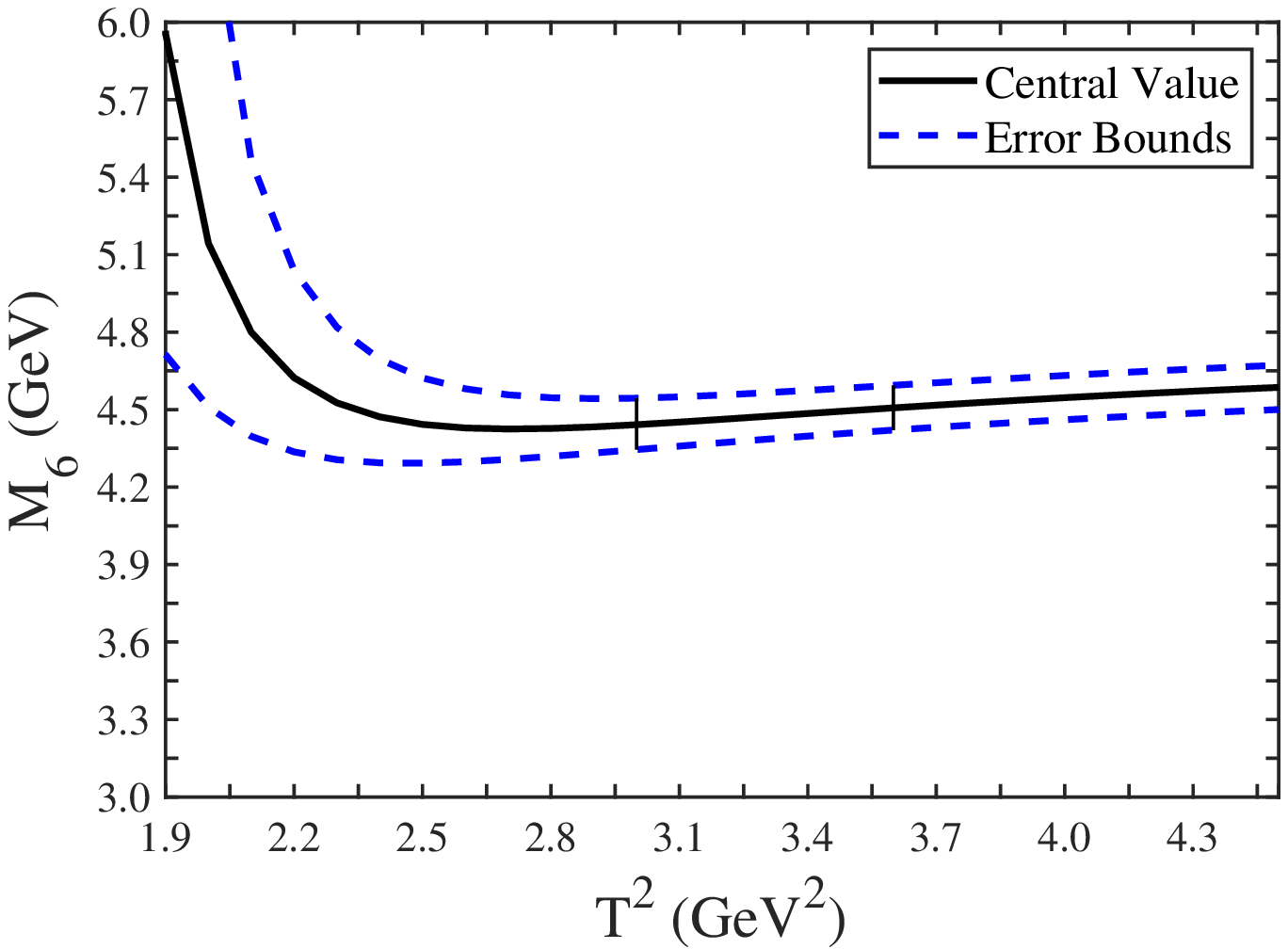}
 \includegraphics[totalheight=5cm,width=7cm]{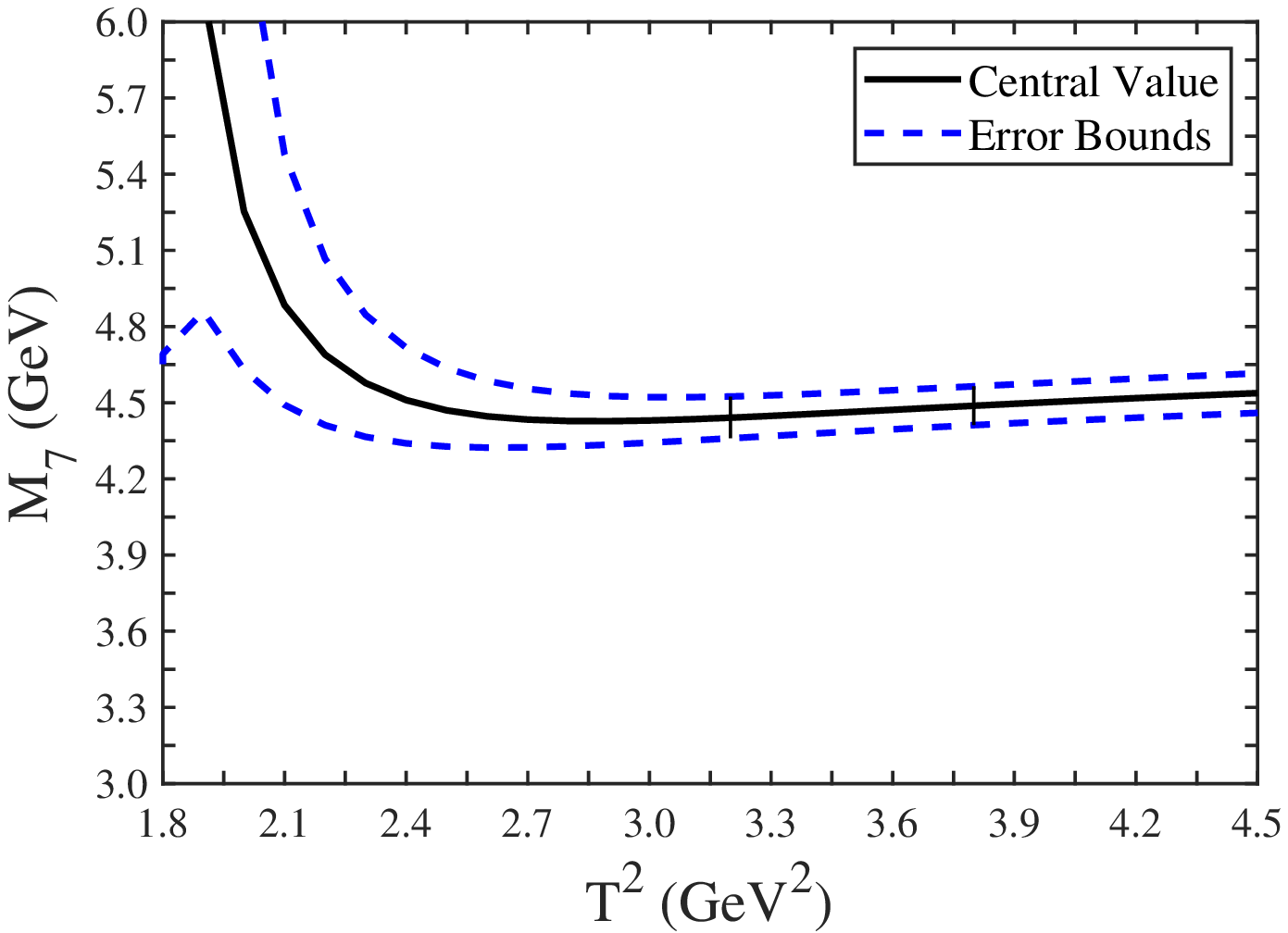}
 \includegraphics[totalheight=5cm,width=7cm]{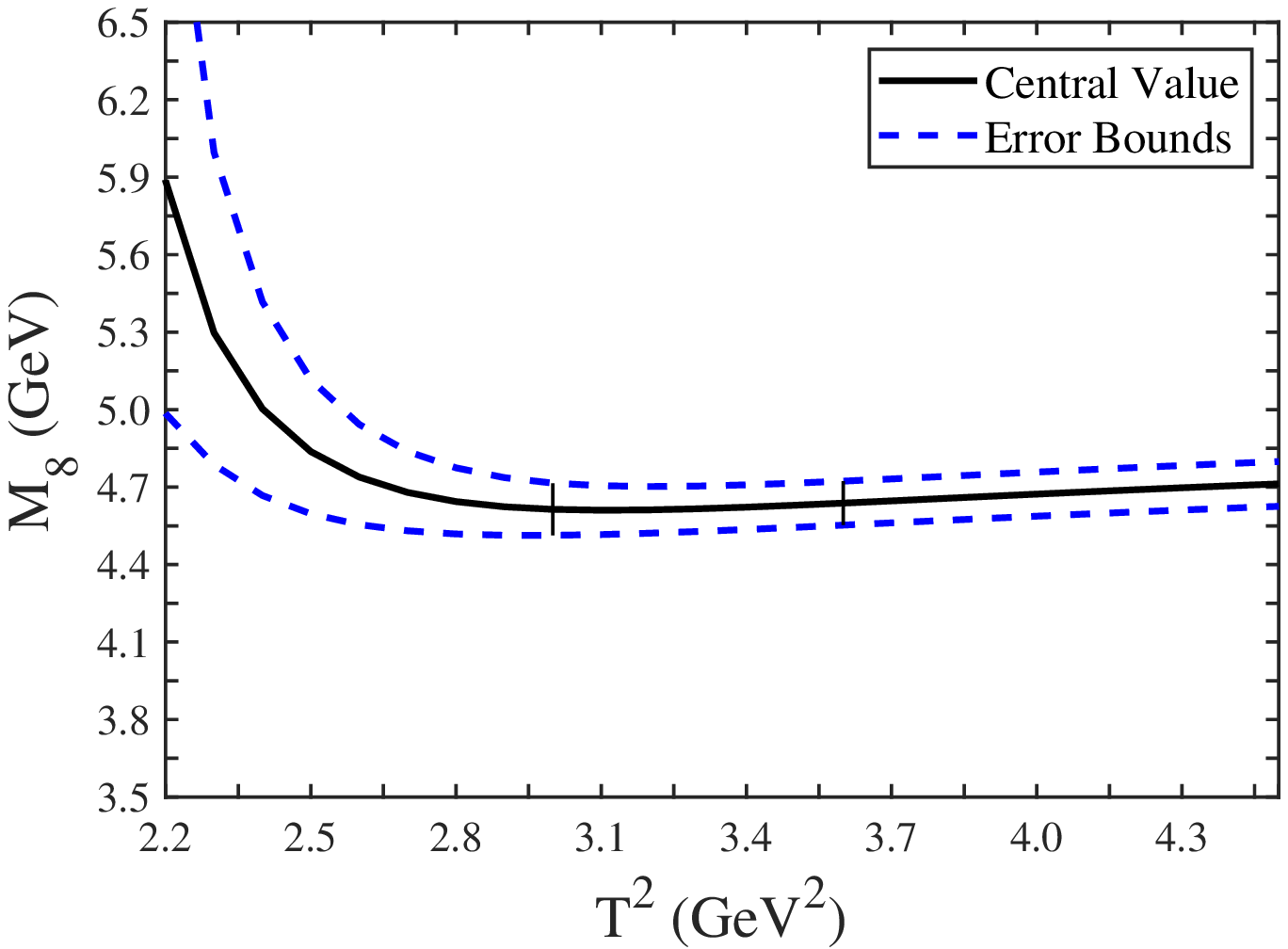}
 \caption{ The $M-T^2$ curves, where $M_i(i=1,2,\cdot\cdot\cdot8)$ denote the masses of the $\bar{D}\Sigma_c$ with
 $I=\frac{1}{2}$, $\bar{D}\Sigma_c$ with $I=\frac{3}{2}$, $\bar{D}\Sigma_c^*$ with $I=\frac{1}{2}$, $\bar{D}\Sigma_c^*$ with
 $I=\frac{3}{2}$, $\bar{D}^*\Sigma_c$ with $I=\frac{1}{2}$, $\bar{D}^*\Sigma_c$ with $I=\frac{3}{2}$, $\bar{D}^*\Sigma_c^*$
 with $I=\frac{1}{2}$ and $\bar{D}^*\Sigma_c^*$ with $I=\frac{3}{2}$, respectively.}\label{mass-fig}
\end{figure}

We calculate the uncertainties of the  masses and pole residues  according to the standard error analysis formula,
 the numerical results of the masses and pole residues are shown in the Table \ref{BorelP-mass-residue} (also the Fig.\ref{mass-fig}).

From Table \ref{BorelP-mass-residue}, we can see that the central value of the extracted mass of the $\bar{D}\Sigma_c$ pentaquark state with the quantum numbers $IJ^P=\frac{1}{2}\frac{1}{2}^-$  is $4.31$ $\rm{GeV}$, it is only about $10$ $\rm{MeV}$ below the $\bar{D}^0\Sigma_c^+$ threshold, so we can assign this state as the $P_c(4312)$ naturally. For the $\bar{D}\Sigma_c$ pentaquark state with the quantum numbers $IJ^P=\frac{3}{2}\frac{1}{2}^-$, the central value of the mass is $4.33$ $\rm{GeV}$, we find it is about $10$ $\rm{MeV}$ above the  $\bar{D}^-\Sigma_c^{++}$ threshold,  so  we can assign this one as the $\bar{D}\Sigma_c$ resonance state, the isospin cousin of the $P_c(4312)$.

In a similar way, according to the numerical results of the extracted masses, we have very good reasons  to assign the  $P_c(4380)$, $P_c(4440)$ and $P_c(4457)$ as the  $\bar{D}\Sigma_c^*$, $\bar{D}^*\Sigma_c$ and $\bar{D}^*\Sigma_c^*$ pentaquark states with the quantum numbers  $IJ^P=\frac{1}{2}\frac{3}{2}^-$,  $\frac{1}{2}\frac{3}{2}^-$ and   $\frac{1}{2}\frac{5}{2}^-$, respectively. For the color singlet-singlet type pentaquark states (resonances) $\bar{D}\Sigma_c^*$ with $IJ^P=\frac{3}{2}\frac{3}{2}^-$, $\bar{D}^*\Sigma_c$ with $IJ^P=\frac{3}{2}\frac{3}{2}^-$ and $\bar{D}^*\Sigma_c^*$ with $IJ^P=\frac{3}{2}\frac{5}{2}^-$, the central values of the extracted masses are about $20$ $\rm{MeV}$, $10$ $\rm{MeV}$ and $90$ $\rm{MeV}$ above the corresponding meson-baryon thresholds, respectively.

If we choose the same input parameters, the color singlet-singlet type pentaquark states with the isospin $I=\frac{3}{2}$ have slightly larger masses than the corresponding pentaquarks with the isospin $I=\frac{1}{2}$.  In calculations,  we observe that the masses and pole residues increase monotonously  with the increase of the continuum threshold parameters, we determine the continuum threshold parameters $s_0$ by adopting the  uniform constraints, such as the continuum thresholds  $\sqrt{s_0}= M_{-}+0.65 \pm0.1 \,\rm{GeV}$, pole contributions $(40\sim 65)\%$ and  intervals   $T^2_{max}-T^2_{min}=0.6\,\rm{GeV}^2$ to acquire reliable predictions, where the  $T^2_{max}$ and $T^2_{min}$ stand for the maximum and minimum values of the Borel parameters, respectively.

It is clearly that the $P_c(4312)$, $P_c(4380)$, $P_c(4440)$ and $P_c(4457)$ can be assigned to be  the  $\bar{D}\Sigma_c$, $\bar{D}\Sigma_c^*$, $\bar{D}^*\Sigma_c$ and $\bar{D}^*\Sigma_c^*$ pentaquark   states with the isospin $I=\frac{1}{2}$, since the two-body strong decays $P_c \to J/\psi p$ conserve isospin. If  the assignments are robust, there exist four slightly higher  pentaquark   states $\bar{D}\Sigma_c$, $\bar{D}\Sigma_c^*$, $\bar{D}^*\Sigma_c$ and $\bar{D}^*\Sigma_c^*$ with the isospin $I=\frac{3}{2}$, we can search for the four resonances in the $J/\psi \Delta$ invariant mass spectrum, as the two-body strong decays $P_c \to J/\psi \Delta$ also conserve isospin, the $J/\psi$, $p$ and $\Delta$ have the isospins $I=0$, $\frac{1}{2}$ and $\frac{3}{2}$, respectively.  If the four resonances are observed one day, we can obtain additional proofs for the color singlet-singlet type pentaquark assignments, and shed light on the nature of the $P_c$ states and dynamics of the low energy QCD.

In this work, we construct the local color singlet-singlet type five-quark currents with the definite isospins, which couple potentially to the  color singlet-singlet type hidden-charm pentaquark states rather than to the meson-baryon scattering states or thresholds, the thresholds in Table \ref{BorelP-mass-residue} are taken from Particle Data Group \cite{PDG},  as the traditional charmed mesons and baryons are spatial extended objects and have average  spatial sizes $\sqrt{\langle r^2\rangle}\approx 0.5\,\rm{fm}$ and $0.5\sim 0.8\, \rm{fm}$, respectively \cite{WZG-penta-mole-CPC,WZG-X4140-X4685}.
Therefore, the loosely bound molecular states, meson-baryon scattering states or thresholds have spatial extensions larger than $1\,\rm{fm}$, which is too large to be interpolated by the local currents.   In the local limit $r \to 0$,  in such small spatial separations, the  $\bar{c}q$ meson
and  $cq q^\prime$ baryon lose themselves and merge into color singlet-singlet type pentaquark states. The scenario of the color singlet-singlet type pentaquark states in the QCD sum rules is quite different from other theoretical methods. In the QCD sum rules, there are two color-singlet clusters, which have the same quantum numbers as the physical states  $\bar{D}^{(*)}$ and $\Sigma_c^{(*)}$, respectively, but they are not the physical states, and we carry out the operator product expansion at the quark-gluon level at the QCD side, and can only distinguish the short distance and long distance contributions,  no hadronic degrees  of freedoms are needed. In fact, we can  abandon the conception "molecular states" in the QCD sum rules, we just investigate  the color singlet-singlet type pentaquark states, which have masses near the meson-baryon thresholds.

 While in the one-pion exchange potential model \cite{Hosaka-OPEP} and heavy-quark spin symmetry model \cite{GengLS-HQSS}, there are physical charmed meson and baryons states. In the one-pion exchange potential model, the short range interaction by the coupling to the 5-quark-core states plays a major role in determining of
the ordering of the multiplet states, while the long range force of the pion tensor force does in producing the decay
widths \cite{Hosaka-OPEP}. In the heavy-quark spin symmetry model, the pentaquark-like resonances can be naturally accommodated
in a contact-range effective field theory description that incorporates the heavy-quark spin symmetry  \cite{GengLS-HQSS}.

The $P_c(4380)$ observed in the six-dimensional amplitude analysis  obsolete in the updated analysis \cite{RAaij2}, which weakens the previously reported evidence for the $P_c(4380)$, but does not contradict its existence, as the one-dimensional analysis is not sensitive to wide $P_c$
 states. Whether or not there exist $P_c(4380)$-like wide pentaquark candidates,  a six-dimensional amplitude analysis of the
$\Lambda_b^0 \to J/\psi p K^- $ decays in the future could be able to answer  the question. Our calculations just indicate that there exists  a color singlet-singlet type pentaquark  candidate with the mass about $4.38\,\rm{GeV}$, and it is not necessary to be the $P_c(4380)$.

In Ref.\cite{WZG-ZHJX-Zc3900}, we  assign the $Z_c^\pm(3900)$ as  the diquark-antidiquark  type   tetraquark  state with the quantum numbers $J^{PC}=1^{+-}$, and study  the hadronic coupling  constants in its two-body strong decays with the QCD sum rules based on the rigorous current-hadron duality,  and obtain  satisfactory total width to match to the experimental data.  We can  explore the two-body strong decays of the color singlet-singlet type pentaquark states  based on the rigorous current-hadron duality, and get  the branching fractions, which can be confronted with  the experimental data in the future to assign the color singlet-singlet type pentaquark states in more reasonable foundations.

\begin{sidewaystable}[thp]
\begin{center}
\begin{tabular}{|c|c|c|c|c|c|c|c|c|c|c|c|c|}\hline\hline
                         &$IJ^P$                        &$T^2({\rm GeV}^2)$   &$\sqrt{s_0}({\rm GeV})$    &$\mu ({\rm GeV})$   &$\rm PC$  &$M({\rm GeV})$  &$\lambda(10^{-3}{\rm GeV}^6)$     & Assignments  &Thresholds (MeV)\\ \hline

$\bar{D}\Sigma_c$        &$\frac{1}{2}\frac{1}{2}^-$    &$3.2-3.8$  &$5.00\pm0.10$  &$2.2$   &$(42-60)\% $ &$4.31^{+0.07}_{-0.07}$  &$3.25^{+0.43}_{-0.41}$   &$P_c(4312)$  &$4321$ \\ \hline
$\bar{D}\Sigma_c$        &$\frac{3}{2}\frac{1}{2}^-$    &$2.8-3.4$  &$4.98\pm0.10$  &$2.2$   &$(44-65)\% $ &$4.33^{+0.09}_{-0.08}$  &$1.97^{+0.28}_{-0.26}$   &resonance  &$4321$ \\ \hline

$\bar{D}\Sigma_c^*$      &$\frac{1}{2}\frac{3}{2}^-$    &$3.3-3.9$  &$5.06\pm0.10$  &$2.3$   &$(42-60)\% $ &$4.38^{+0.07}_{-0.07}$  &$1.97^{+0.26}_{-0.24}$   &$P_c(4380)$  &$4385$ \\ \hline
$\bar{D}\Sigma_c^*$      &$\frac{3}{2}\frac{3}{2}^-$    &$2.9-3.5$  &$5.03\pm0.10$  &$2.4$   &$(44-64)\% $ &$4.41^{+0.08}_{-0.08}$  &$1.24^{+0.17}_{-0.16}$   &resonance  &$4385$\\ \hline

$\bar{D}^*\Sigma_c$      &$\frac{1}{2}\frac{3}{2}^-$    &$3.3-3.9$  &$5.12\pm0.10$  &$2.5$   &$(42-60)\% $ &$4.44^{+0.07}_{-0.08}$  &$3.60^{+0.47}_{-0.44}$   &$P_c(4440)$ &$4462$\\ \hline
$\bar{D}^*\Sigma_c$      &$\frac{3}{2}\frac{3}{2}^-$    &$3.0-3.6$  &$5.10\pm0.10$  &$2.5$   &$(41-61)\% $ &$4.47^{+0.09}_{-0.09}$  &$2.31^{+0.33}_{-0.31}$   &resonance\ &$4462$\\ \hline

$\bar{D}^*\Sigma_c^*$    &$\frac{1}{2}\frac{5}{2}^-$    &$3.2-3.8$  &$5.08\pm0.10$  &$2.5$   &$(43-60)\% $ &$4.46^{+0.08}_{-0.08}$  &$4.05^{+0.54}_{-0.50}$   &$P_c(4457)$  &$4527$\\ \hline
$\bar{D}^*\Sigma_c^*$    &$\frac{3}{2}\frac{5}{2}^-$    &$3.0-3.6$  &$5.24\pm0.10$  &$2.8$   &$(42-61)\% $ &$4.62^{+0.09}_{-0.09}$  &$2.40^{+0.37}_{-0.35}$   &resonance &$4527$\\
\hline\hline
\end{tabular}
\end{center}
\caption{ The Borel parameters, continuum threshold parameters, energy scales, pole contributions, masses, pole residues and assignments for the eight color singlet-singlet type  pentaquark states, where the thresholds denote the corresponding thresholds of the  meson-baryon scattering states. }\label{BorelP-mass-residue}
\end{sidewaystable}

\section{Conclusions}
In the present work, we distinguish the isospins of the color singlet-singlet type pentaquark states   and construct the color singlet-singlet type five-quark currents with the isospins $(I,I_3)=(\frac{1}{2},\frac{1}{2})$ and $(\frac{3}{2},\frac{1}{2})$ unambiguously to  explore their properties with the QCD sum rules for the first time. In order to obtain accurate numerical results, we consider the vacuum condensates  up to dimension $13$ in a consistent way. Based on the extracted pentaquark masses from the Borel windows, we assign the $\bar{D}\Sigma_c$, $\bar{D}\Sigma_c^*$, $\bar{D}^*\Sigma_c$ and $\bar{D}^*\Sigma_c^*$ pentaquark  states with the isospin $I=\frac{1}{2}$ to be the $P_c(4312)$, $P_c(4380)$, $P_c(4440)$ and $P_c(4457)$, respectively, see Table \ref{BorelP-mass-residue}. Furthermore, the present calculations indicate that there also exist four  slightly higher  pentaquark  states $\bar{D}\Sigma_c$, $\bar{D}\Sigma_c^*$, $\bar{D}^*\Sigma_c$ and $\bar{D}^*\Sigma_c^*$ with the isospin $I=\frac{3}{2}$, which lie slightly above the thresholds of the corresponding meson-baryon pairs  $\bar{D}\Sigma_c$, $\bar{D}\Sigma_c^*$, $\bar{D}^*\Sigma_c$ and $\bar{D}^*\Sigma_c^*$, respectively. We can search for the four resonances in the $J/\psi \Delta$ invariant mass spectrum, which can lead to additional proofs for the color singlet-singlet type pentaquark  assignments, and shed light on the nature of the $P_c$ states and dynamics of the low energy QCD.

\section*{Data Availability}

 All data included in this manuscript are available upon request by contacting with the correspond-
ing authors.

\section*{Conflicts of Interest}

 The authors declare that they have no conflicts of interest.

\section*{Acknowledgements}
This work is supported by National Natural Science Foundation, Grant Number 12175068 and Youth Foundation of NCEPU, Grant Number 93209703.

\end{document}